\documentstyle[aps,prl,amsfonts,amssymb,floats,twocolumn,epsfig]{revtex}
\begin{document}
\draft
\twocolumn[\hsize\textwidth\columnwidth\hsize\csname
@twocolumnfalse\endcsname
\title{Effects of Vortex Pinning and Thermal Fluctuations on the Josephson Plasma
Resonance in Tl$_2$Ba$_2$CaCu$_2$O$_8$ and YBa$_2$Cu$_3$O$_{6.5}$}
\author{Diana Duli\'c$^1$, S. J. Hak$^1$,  D. van der Marel$^1$, W. N. Hardy$^2$,
A. E. Koshelev$^3$, Ruixing Liang$^2$, D. A. Bonn$^2$, B. A.
Willemsen$^4$}
\address{$^1$Laboratory of Solid State Physics, Materials Science Centre,
Nijenborgh 4, 9747 AG Groningen, The Netherlands}
\address{$^2$ Department of Physics and Astronomy, University of British
Columbia, Vancouver, BC, V6T 1Z1, Canada}
\address{$^3$ Material Science Division, Argonne National Laboratory,
Argonne, IL 60439}
\address{$^4$ Superconductor Technologies Inc., Santa Barbara,CA
93111-2310}
\date{\today}
\maketitle
\begin{abstract}
Infrared spectroscopy has been used to investigate the temperature
dependence and $c$-axis magnetic field dependence of the Josephson
plasma resonance in optimally doped Tl$_2$Ba$_2$CaCu$_2$O$_8$ thin
films, and underdoped YBa$_2$Cu$_3$O$_{6.5}$ ortho - II single
crystals. The resonances are directly related to the $c$-axis
penetration depths, yielding low temperature, zero fields values
of about 20 $\mu$m, and 7 $\mu$m respectively. While the temperature
dependencies of the resonances in the two compounds are very
similar, the magnetic field dependence in YBCO is much weaker. We
attribute this weak magnetic field dependence to the lower
anisotropy of YBCO, and discuss the observed behaviours in terms
of thermal fluctuations and pinning of pancake vortices, valid for highly
anisotropic layered superconductors.
\end{abstract}
\vskip2pc] \narrowtext
%
%
\begin{figure}[h]
 \centerline{\epsfig{figure=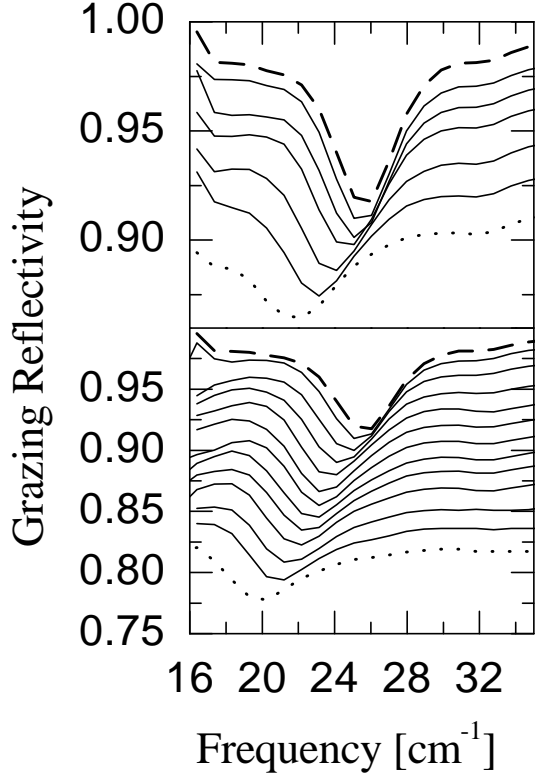,width=7cm,clip=}}
 \caption{Top panel: Grazing incidemce reflectivity of Tl2212
 without external magnetic field at (from top to bottom) 10, 20, 
 30, 40, 50, 60, and 70 K.
 Bottom panel:Field cooled grazing incidence reflectivity of Tl2212 
 at 4 K in an external magnetic field of (from top to bottom) 
 0, 0.03, 0.06, 0.11, 0.17, 0.21, 0.23, 0.26, 0.29, 0.32, 0.36,
 0.4, and 0.46 Tesla, perpendicular to the surface. 
 Except for the zero field data, all curves have been given 
 vertical offsets.}
 \label{tlspect}
 \end{figure}
The Josephson plasma resonance (JPR) phenomenon\cite{pwa,dahm,gubankov} provides direct
information about the Josephson coupling between the CuO$_2$
planes in the highly anisotropic high T$_c$ superconductors\cite{marel}.
Studies of the temperature and magnetic field dependence of the JPR
provide useful insights into the pinning and thermal fluctuations of
vortices in the cuprate family\cite{tsui,matsuda1,matsuda,bulaevskiib,bulaevskiia,alexc}.
It is natural, to anticipate unusual magnetic field behavior in
these highly anistropic materials. Indeed, just as in any conventional
type-II superconductor, if a DC magnetic field is applied
perpendicular to the 2D planes it penetrates the sample in the
form of vortices which, when the anisotropy is large, are not
vortex lines but pancake vortices. For low magnetic fields and low
temperatures, these pancake vortices {\em may} form a nearly
perfect vortex lattice. Under these conditions there should be no
magnetic field dependence of the JPR since
$\omega_{p}^{2}(B,T)=\omega_{p}^{2}(0)\langle\cos(\phi_{n,n+1})(r)\rangle$,
where $\omega_{p}^{2}(0)$ is the plasma frequency in zero field,
and $\phi_{n,n+1}(r)$ is the gauge-invariant phase difference
between the neighboring layers $n$ and $n+1$\cite{bulaevskiib}. In
zero field, $\phi_{n}(r)$ is independent of $r$ and independent of
$n$ for zero $c$-axis current. In non-zero field, $\phi_{n}(r)$
becomes position dependent but $\phi_{n,n+1}(r)$ is everywhere
zero, provided the vortex cores are lined up, as in a perfect
lattice. However, due to the effects of pinning (at low
temperatures), and thermal fluctuations (at higher temperatures)
pancake vortices do not in general line up from layer to layer.
Both regimes have been extensively studied theoretically  by
Bulaevskii and one of us (AEK)\cite{bulaevskiia,alexc,alexa}, but
there is very little data on the JPR in the low magnetic field
regime\cite{matsuda}. In this paper we present our studies of the
temperature and field dependence of the JPR in two members of the
high T$_c$ family with differing anisotropies.

Two types of reflectivity spectra were measured, each optimized to
the type of sample: First normal incidence reflectivity spectra
were recorded of a mosaic made of six $ac$-oriented single
crystals of YBa$_2$Cu$_3$O$_{6.5}$ with ortho-II oxygen ordering,
and a superconducting transition temperature of 58K. The mosaic
was held together with epoxy and had an area of 6mm$^2$. Second,
grazing incidence reflectivity measurements were made on
epitaxially grown Tl$_2$Ba$_2$CaCu$_2$O$_8$ thin films, with a
superconducting transition temperature of 98K. Dimensions of the
films in the $ab$ plane were 64 mm$^2$, and the thickness was
600-700 nm. Details of the sample preparation and growth methods
have been reported elswhere\cite{willemsen,eddy}. Here we report
data in the far infrared FIR region (17-700 cm$^{-1}$) obtained
using a Fourier transform spectrometer. A grazing angle of
80$^{\circ}$ was chosen to probe predominantly the $c$-axis
response for $p$-polarization of the light. Absolute
reflectivities were obtained by calibrating the reflectivities
against the reflectivity of a gold film deposited {\em in situ} on
the sample.

The reflection experiments were carried out in magnetic fields
oriented along the $c$-axis, which could be varied continuously
between 0 and 0.8T using a sliding permanent magnet, and calibrated
using a GaAs 2D electron gas placed in the sample holder block,
1 mm underneath the sample. Details of the magnet-cryostat design 
are described elsewhere\cite{dulic}.

\begin{figure}
\centerline{\epsfig{figure=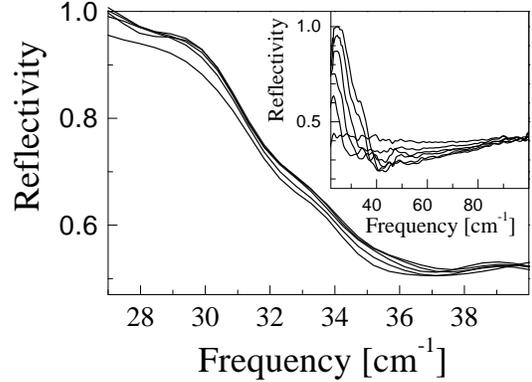,width=7cm,clip=}}
 \caption{Field-cooled normal incidence reflectivity of YBCO.
  at 4K in an external magnetic field of (from top to bottom) 
 0, 0.2, 0.4, 0.6, and 0.8 Tesla, perpendicular to the surface. 
 Inset: Reflectivity of YBCO without external magnetic field at 
 (from top to bottom) 4, 30, 40, 50, and 70 K}
 \label{yfc}
\end{figure}

\begin{figure}
 \centerline{\epsfig{figure=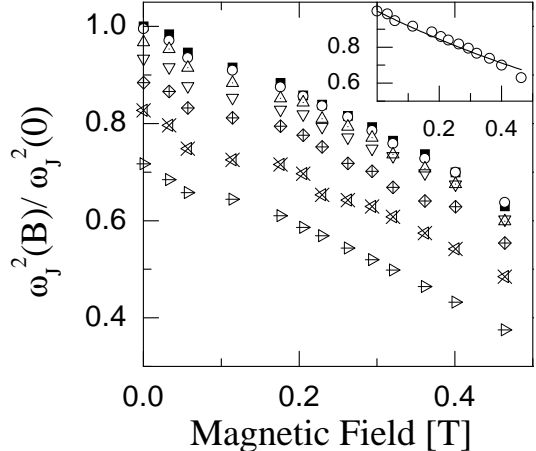,width=7cm,clip=}}
 \caption{Magnetic field dependence of $\omega(B,T)^{2}$
 normalized to the value at zero field at 4K (field-cooled) in
 Tl2212 for temperatures from top to bottom:
 10K, 20K, 30K, 40K, 50K, 60K, 70K. Inset:  4K data, with a
 fit to Eq.\ref{eq:kos1}, using $B_w = 1.40 T$.} 
 \label{magcomb}
\end{figure}

In Fig.~\ref{tlspect} we show the grazing incidence reflectivity
of Tl2212 for a range of temperatures in zero field, and for various 
magnetic fields at 4 K under field cooled (FC) conditions:
The sample temperature was first increased above the phase 
transition, next the field  was set at the desired value, the 
sample temperature was lowered to 4 K, and the spectra were 
recorded. This sequence was then repeated for each external field value.
In Fig.~\ref{yfc} we present the FC normal incidence reflectivity
of YBCO, with the electric field along the $c$-axis in the
frequency region from 18-100 cm$^{-1}$ for various magnetic 
fields at 4K. In the figures we show
only the low frequency part of the spectra since the only features
in the spectra below 50 cm$^{-1}$ correspond to the JPR.
However, for the spectral analysis and accurate determination of
the plasma resonance frequency, the entire measured spectral range
was used. Fig.\ref{tlspect} reveals that the JPR shifts down  
quasi-linearly as a function of magnetic field for all temperatures
(about 10 cm$^{-1}$ per Tesla). 
The gradual redshift without a change of amplitude upon increasing the 
magnetic field, indicates that the resonance is a Josephson plasma resonance for 
all fields considered, and not a vortex mode\cite{sonin}.  
The FC magnetic field dependence of 
the JPR of Tl2212 is displayed in Fig. \ref{magcomb} for each temperature. 
For YBCO the JPR has a much lower field dependence of about 
$0.5$ cm$^{-1}$ per Tesla (see Fig. \ref{yfc}). 
%
%

The $c$-axis penetration depths extracted from the zero field plasma 
frequency at 4K are 19.8 $\mu$m in Tl2212 and 7.1 $\mu$m in YBCO,
using $\epsilon_S=9.1$\cite{dulic99} and $\epsilon_S=28$\cite{homes} respectively.
Previously we have addressed the issue of the temperature
dependence of the resonance\cite{dulic99} in Tl2212 and Tl2201, from
which we obtained the temperature dependence of the $c$-axis penetration 
depth, $\lambda_c^2(0)/\lambda_c^2(T) = 1 - (T/T_0)^{\eta}$ with
$\eta\sim 2$. From the present data on YBCO we obtained the same 
temperature dependence.

\begin{figure}
 \centerline{\epsfig{figure=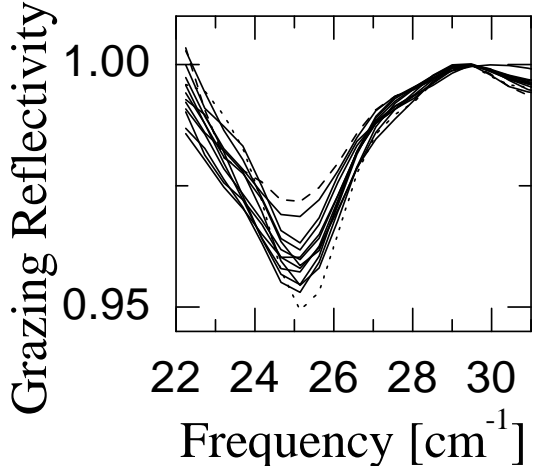,width=7cm,clip=}}
 \caption{Zero-field cooled grazing incidence reflectivity  of 
 Tl2212 at 10K in an external magnetic field of (from top to 
 bottom) 0.46, 0.39, 0.32, 0.25, 0.18, 0.12, 0.06 and 0 Tesla, 
 perpendicular to the surface}
 \label{tzfc}
\end{figure}

In addition we measured reflectivity data under zero field cooled 
(ZFC) conditions. These data are displayed in Fig. \ref{tzfc}, showing 
that under ZFC conditions the resonance stays at the same frequency, 
whereas the {\em intensity} of the
resonance drops gradually upon raising the field. The
intensity also exhibits hysteresis during cycling of the
magnetic field (not shown). Briefly, this behaviour is caused by the self
screening of the sample: As the field is applied perpendicular to
the film, the outer parts of the film are penetrated by the field,
while the central part of the film remains at $B=0$. Assuming that
the JPR is absent from the field penetrated region of
the film, the spectra are just linear combinations of the
zero field spectrum and the featureless high field spectrum.
One can make this quantitative using the result of Mikheenko
and Kuzovlev\cite{mikheenko} for the distance $a$ from the center
of the disk that is penetrated by the magnetic field:
$a=R/\cosh(H/(2\pi j_{c}^{ab}d))$, where $R$ is the radius of the disk,
$H$ is the applied field, $j_{c}^{ab}$
is the $ab$-plane critical current and $d$ is the thickness of the
film. From comparison to Fig. \ref{tzfc} we obtained 
$j_{c}^{ab}=6.7*10^{10}$A/m$^{2}$, which is a 
reasonable value for a high quality film.

%

The resonance is due to the zero-crossing
of Re$\epsilon_{c}(\omega)$, for which we adopt the 
two-fluid expression appropriate for the superconducting state
\begin{equation}
 \epsilon_{c}(\omega)=\epsilon_{S}(\omega)
  - \frac{c^2}{\lambda_c^{2}(B,T)\omega^{2}}
  + \frac{4\pi i \sigma_{qp}(\omega)}{\omega}
 \label{eq:difun1}
\end{equation}
where $\lambda_c$ is the
c-axis penetration depth, and $1/\lambda_c(B,T)^2$ represents the
superfluid density appropriate for the $c$-direction. The
pole-strength of the second (London) term, coming from the
condensate provides directly, within a Josephson coupled layer
model, the critical current in the $c$-direction. Oscillations of
the condensate are electromagnetically coupled to interband
transitions and lattice vibrations, which are represented by the
dielectric function $\epsilon_{S}$. The third term arises from
dissipative currents due to quasi-particles, and determines the 
width of the plasma resonance in our spectra\cite{dulic99}.
The condition for propagation of longitudinal modes in the medium along 
the c-direction in the long wavelength limit, is that
$\epsilon_{c}(\omega)=0$, which occurs at the JPR frequency
$\omega_{J}=c/\{\lambda_c\sqrt{\mbox{Re}\epsilon_{S}(\omega_{J})}\}$.
By fitting Eq. \ref{eq:difun1} to the measured spectra, using the full 
Fresnel expression for the grazing/normal incidence reflectivity
of uniaxial crystals\cite{dulic99}, we obtain accurate values of 
the $c$-axis superfluid density, $\lambda_c^2(0)/\lambda_c^2(B,T)$, 
as a function of temperature and magnetic field.
  
As mentioned above, a DC magnetic field applied along the $c$-axis
penetrates the cuprate superconductors in the form of pancake
vortices. For sufficiently low fields and temperatures they form a
vortex lattice, which for an ideal material would be composed of straight
lines along the $c$-axis, and as a consequence have no influence on
the inter-layer Josephson coupling. If the coupling between the
layers is very weak, pinning and thermal fluctuations may be
sufficient to destroy the alignment of the pancake vortices along
the field direction. The resulting spatial fluctuations of the
relative phases between the planes cause a suppression of the
Josephson coupling between the layers. Below the irreversibility
line, the vortex phase diagram is very
rich\cite{matsuda,kes}, and similar richness
is required of theories aimed at explaining all parts of the
diagram. In the low temperature regime the main cause of
misalignment of the pancakes is pinning due to impurities in the
materials. At higher temperatures, thermal fluctuations take over
and become the main cause for suppression of the Josephson
current. Suppression of the Josepson coupling due to a magnetic
field applied along the $c$-axis in the vortex lattice state was
extensively studied theoretically by one of us (AEK) in both the
low temperature (where the pinning is the most important
contribution), and the high temperature (where thermal
fluctuations play the most important role) regimes. In
Ref.\cite{alexa} a model was developed where the influence of weak
pinning on the Josephson coupling between layers was considered. A
characteristic "decoupling" field $B_w$ was introduced, above which the JPR
frequency saturates at a certain value. The vortex lines are destroyed by 
pinning, and $B_w$ is determined by the wandering length of the vortex lines. 
The following expression was derived for the effective Josephson coupling
\begin{equation}
 E_J^{eff}-E_J=-\frac{E_JB}{B_w}
 \label{eq:kos1}
\end{equation}
Here $E_J$ is Josephson energy per unit area, which is related
to the $c$-axis penetration depth via
$E_J= \Phi_0^2 / ( 16 \pi^3 \lambda_c^2 d)$,
where $\Phi_0$ is the elementary flux quantum, and 
$d$ is the interlayer distance.  With this expression we 
can calculate the Josephson energies per unit area with the result
$E_J=1.6\cdot 10^{-4} erg/cm^{2}$ for Tl2212, and
$E_J=1.5\cdot 10^{-3} erg/cm^{2}$ in YBCO. In the inset of
Fig.\ref{magcomb} we show the fit to Eq.\ref{eq:kos1} at 4K from
which we obtain values for $B_w=1.40\pm 0.02 T$ in Tl2212. In 
YBCO the field dependence is much weaker (Fig. \ref{yfc}),
corresponding to $B_w=30 \pm 10 T$. Now, we can pose the question why YBCO
has a weaker field dependence than Tl2212. From the discussion above
it is clear that it can either be due to the higher
purity of the sample or to the more three dimensional nature of
YBCO. The theoretical prediction for $B_w\sim E_J\Phi_{0}/U_p$
from Ref.\cite{alexa}, where $U_p$ represents the pinning
potentional can help us resolve this problem. The
ratios of the 'decoupling' fields for Tl2212 and YBCO should scale as
$B_w^{Tl}/B_w^{Y}=(E_J^{Tl}/(E_J^{Y}) \cdot (U_p^{Y}/U_p^{Tl})$.
Since we have already determined
the Josephson coupling constants $E_J$ and the decoupling fields
$B_w$, we have sufficient experimental information to estimate the 
ratio of the pinning potentials,providing $U_p^{Y}/U_p^{Tl} \approx 0.44$. 
In the strongly anisotropic material Bi2212
$\lambda_c(0) = 110 \mu$m \cite{matsuda}, corresponding to
$E_J=6.0\cdot10^{-6} erg/cm^{2}$, while B$_w \approx 0.12$ Tesla\cite{matsuda}. 
In the same way as for YBCO we can estimate, that $U_p^{Bi}/ U_p^{Tl} =0.43$. 
The ratio's of the 'decoupling' fields of these
three compounds are $B_w^{Bi}:B_w^{Tl}:B_w^{Y}=1:12:250$. 
The ratio's of the pinning potentials are 
$U_p^{Bi}:U_p^{Tl}:U_p^{Y}=1:2:1$, which does not correlate with 
B$_w$ in Bi2212, Tl2212, and YBCO. On the other hand, the relative values of the
Josephson coupling energies are $E_J^{Bi}:E_J^{Tl}:E_J^{Y}=1:27:250$, from
which we can conclude that the spread of $B_w$'s over two orders of magnitude 
is a consequence of the strong material dependence of the interlayer 
Josephson coupling constant. 

%

\begin{figure}
\centerline{\epsfig{figure=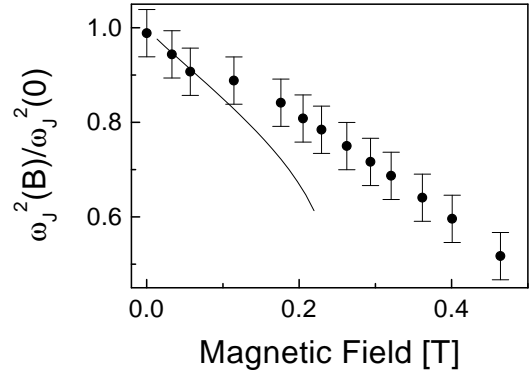,width=7cm,clip=}}
 \caption{
  Comparison between the experimental values of
  $\omega(B,T)^{2}/\omega(0,T)^{2}$ of Tl2212 at 70 K
  and the theoretical prediction based on Ref. 10
  with the the parameters $\lambda_{ab} = 245 nm$, 
  and $\lambda_c = 24.2 \mu$m for the
  in-plane and c-axis penetration depth.}
 \label{kosh}
\end{figure}

As temperature is increased towards T$_c$, a cross-over may take place 
to a different type of magnetic field suppresion of the JPR, where the 
pancake vortices become progressively misaligned due to thermal fluctuations. In 
Ref. \cite{alexc} it was shown for low fields in the thermal fluctuation regime, that
$\omega_J^2(0,T)-\omega_J^2(B,T) \propto B$ in the low field region, with a gradual
downward curvature at higher field values, similar to our experimental
data\cite{alexc}. The numerical calculation 
for the present case is shown in Fig. \ref{kosh}, for T=70 K, using 
the experimental values for $\lambda_{ab}(T)= 245$nm\cite{willemsen,jacobs,lee}, and 
$\lambda_c(T)=24.2\mu$m\cite{dulic99} as the (only, experimentally fixed) parameters. 
We see, that, although the field dependence predicted by the thermal fluctuation model 
is close to the observations at 70K, still the experimental field dependence is weaker
than in the calculation. This is probably due to the fact, that there
is still a significant contribution from pinning even at 70 K. This is also born out
by the fact, that in Fig. \ref{magcomb} there is no indication of a sudden change
of the field dependence as a function of temperature. A quantitative theoretical 
framework of these observations would require a unified theoretical model, which
takes into account both the influence of pinning and of thermal fluctuations on the 
Josephson coupling.

%

In conclusion, we have studied the $c$-axis Josephson plasma
resonance in  Tl2212 and YBCO as a function of temperature and
low magnetic field. In the low temperature regime the observed
behaviour of Tl2212 fits very well the model of vortex disorder
due to weak pinning\cite{alexa} while above 50 K, good
quantitative agreement with model calculations
provides evidence for the presence of disorder due to 
thermally fluctuating pancakes\cite{alexc}.
We attribute the weaker magnetic field dependence of the
resonance in YBCO to the lower anisotropy.

We gratefully acknowledge L.N. Bulaevskii for many stimulating
discussions and comments, and P.M. Koenraad at the University of 
Eindhoven for assisting us with the Hall-probe. Work in Argonne was supported by the U.S. DOE,
Officie of Science, under contract No. W-31-109-ENG-38. This investigation was supported 
by the Netherlands Foundation for Fundamental Research on Matter 
(FOM) with financial aid from the Nederlandse Organisatie voor 
Wetenschappelijk Onderzoek (NWO).
\end{document}